\documentclass[aps,showpacs,prl,amssymb,amsmath,twocolumn]{revtex4-2}
\usepackage{graphicx}
\usepackage{amssymb}
\usepackage{amsmath}
\usepackage{amsthm}
\usepackage{bm}
\usepackage{xcolor} 
\usepackage{array}
\usepackage{multirow}
\usepackage{tabularx}
\usepackage{booktabs}
\usepackage{textcomp}
\usepackage{mathtools}
\usepackage{hyperref}
\usepackage{qcircuit}
\usepackage{bbm}
\usepackage{cancel}
\usepackage{comment}
\usepackage{physics}
\usepackage{multirow}

\begin{document}

\title{Quantum Programming of the Satisfiability Problem with Rydberg Atom Graphs}
\author{Seokho Jeong$^1$, Minhyuk Kim$^1$, Minki Hhan$^2$, and Jaewook Ahn$^1$} 
\address{$^1$Department of Physics, Korea Advanced Science and Technology (KAIST), Daejeon 34141, Republic of Korea } 
\address{$^2$ Quantum Universe Center, Korea Institute for Advanced Science (KIAS), Seoul 02455, Republic of Korea}
\date{\today}

\begin{abstract} \noindent
Finding a quantum computing method to solve nondeterministic polynomial time (NP)-complete problems is currently of paramount importance in quantum information science. Here an experiment is presented to demonstrate the use of Rydberg atoms to solve (i.e., to program and obtain the solution of) the satisfiability (3-SAT) problem, which is the prototypical NP-complete problem allowing general programming of all NP problems. Boolean expressions of the 3-SAT problem are programmed with the blockade interactions of Rydberg atom graphs and their many-body ground states are experimentally obtained, to determine the satisfiabilities of the given 3-SAT problem instances quantum mechanically. 
\end{abstract}

\maketitle \noindent
Currently there are considerable efforts being devoted to making a quantum computer~\cite{Arute2019,Ebadi2021,Monroe2021}. One prominent goal is to engineer a quantum system that can formulate quantum algorithms of classically difficult computational problems~\cite{Shor1994,Grover1997}. According to the Cook-Levin theorem~\cite{Cook1971}, an efficient algorithm which can solve a problem in the computational complexity class of non-deterministic polynomial (NP)-complete can be used as a subroutine for the efficient algorithm for all other problems in NP. So, if a quantum computer can solve an NP-complete problem efficiently, all other NP problems can also be efficiently solvable by the polynomial time reduction to the NP-complete problem~\cite{Farhi2001, Dickson2011}. Boolean satisfiability problem (SAT or B-SAT), and the 3-SAT problem that has clauses of at most three literals, are a prototypical NP-complete problem that belongs to the class of NP-complete, i.e., no classical algorithms can efficiently (i.e., in a polynomial time) solve the 3-SAT problem, unless P=NP~\cite{Barahona1982,Arora2009}. 
There are limited physical implementations of the 3-SAT problem, which include an algorithmic conversion to a network-based biocomputation format~\cite{Biocomputation2022}, a quantum circuit approach using Grover's quantum search algorithm in conjunction with David-Putnam-Logemann-Loveland algorithm~\cite{DPLL2020}, and an IBM-Q operation of the Grover's quantum algorithm for the 3-SAT problem~\cite{3SAT_IBMQ2020}. However, these approaches are nonimmune to errors, so it may be worthy for the current noisy-intermediate scale (NISQ) quantum computers to consider the robustness of quantum adiabatic computing.

Of particular relevance in the context of the present paper, the 3-SAT problem is reducible to the maximum independent set (MIS) problem, which is also the NP-complete problem~\cite{VickyChoi2010,Lucas2014}, and the MIS problem is physically implementable with Rydberg atom graphs~\cite{Pichler2018,MHKim2022}. So, in this paper, we introduce a quantum algorithm to formulate the 3-SAT problem with Rydberg atoms; we formulate a quantum experiment to obtain the MIS solution of Rydberg-atom graphs programmed to algorithmically determine a given 3-SAT problem instance, i.e., to evaluate the satisfiability of the 3-SAT instance experimentally.

The 3-SAT problem is to determine whether a given propositional logic formula (Boolean expression), $\Psi(x_1,x_2,\cdots)$, of Boolean variables, $x_1,x_2,\cdots$, is {\it satisfiable} (i.e., there exists a set of Boolean values for the variables satisfying the formula) or {\it unsatisfiable}. The 3-SAT formula is given in the conjunctive normal form~\cite{Hans1999}, i.e, a conjunction of $N_C$ clauses, $\Psi(x_1, x_2, \cdots, x_n)=\bigwedge_{j=1}^{N_C} C_j$, where each clause, $C_j=\ell_{j,1} \vee \ell_{j,2}$ or $\ell_{j,1} \vee \ell_{j,2} \vee \ell_{j,3}$, is a disjunction of at most three literals, $\ell_{j,1}, \ell_{j,2}, \ell_{j,3} \in \left\{x_k,\bar x_k | k=1,\cdots,n \right\}$~\cite{Karp1972,Freeman1979}. The given 3-SAT problem can be reduced to the MIS problem for an MIS graph $G(V,E)$, given by
\begin{eqnarray}
V&=&\left\{(j,k) | \ell_{j,k} \in C_j\right\} \label{V_GM}  \nonumber \\
E&=&E_1 \cup E_2 \label{E_GM} \nonumber \\
E_1&=&\left\{[(j,k_1),(j,k_2)] |k_1 \neq k_2\right\} \label{E1_GM} \nonumber \\
E_2&=&\{ [(j_1,k_1),(j_2,k_2)] | k_1\ \neq k_2, \ell_{j_1,k_1} = \bar \ell_{k_2,k_2} \} \nonumber
\end{eqnarray}
where $V$ is the set of vertices of which an element $(j,k)$ corresponds to the literal $\ell_{j,k}$ in the $j$-th clause $C_j$; $E$ is the set of all edges, the union of two edge sets $E_1$ and $E_2$;  $E_1$ is the set of all intra-clause edges connecting two vertices corresponding to the literals $\ell_{j,k_1}$ and $\ell_{j,k_2}$ in the same clause $C_j$; and
$E_2$ is the set of all inter-clause edges which connect two vertices in different clauses, whose corresponding literals are negation to each other~\cite{VickyChoi2010,Lucas2014}. 

\begin{figure*}[t]
\centerline{\includegraphics[width=0.9\textwidth]{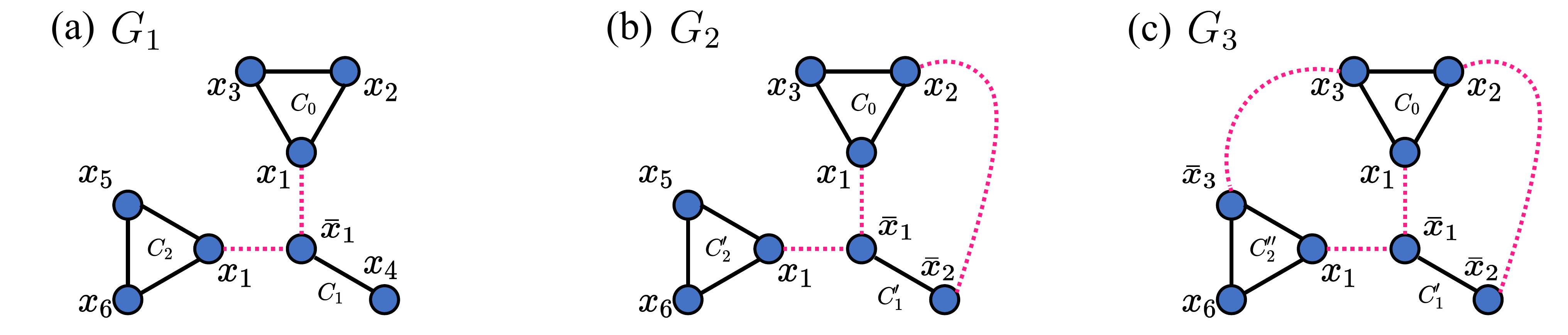}}
\caption{ (a) The MIS graph $G_{1}$ reduced from the 3-SAT instance $\Psi_{1}$ in Eqs.~\eqref{Psi's}, (b) $G_2$ from $\Psi_{2}$, and (c) $G_3$ from $\Psi_{3}$, where vertices represent literals ($x_1, \cdots, x_6$ and negations), solid edges intra-clause logics, and dashed edges the inter-clause logics (between literals and their negations).}
\label{Fig1}
\end{figure*}
Figure~\ref{Fig1} shows examples of MIS graphs obtained with the above reduction algorithm. The first graph, $G_1$ in Fig.~\ref{Fig1}(a) is for a 3-SAT instance, given by 
\begin{subequations} \label{Psi's}
\begin{eqnarray}
\Psi_1(x_1,x_2,x_3,x_4,x_5,x_6) & =& C_0 \wedge C_1 \wedge C_2 \\ 
C_0  &=& x_1 \vee x_2 \vee x_3  \\
C_1  &=& \bar x_1 \vee x_4 \\
C_2  &=& x_1 \vee x_5 \vee x_6
\end{eqnarray} 
\end{subequations}
where the clauses having two or three literals are respectively mapped to the two- or three-vertex subgraphs (of solid edges) and the literal-negation pairs are mapped to the inter-clause edges (of dashed lines). Similarly, we define two more MIS graphs $G_{2}$ and $G_3$ for $\Psi_2 = C_0 \wedge C'_1 \wedge C_2$ and $\Psi_3 = C_0 \wedge C'_1 \wedge C'_2$, with $C'_1 = \bar x_1 \vee \bar x_2$ and $C'_2 = x_1 \vee \bar x_3 \vee x_6$,  as respectively shown in Figs.~\ref{Fig1}(b,c).

To solve the 3-SAT problem quantum mechanically, we perform many-body ground-state searching experiments of Rydberg atoms arranged for the corresponding MIS graphs. The experimental setup is reported elsewhere~\cite{WJLeePRA2019, YSong2021, MHKim2022,AndyByun2022} (see Supplementary for details). 
The Hamiltonian of Rydberg atoms arranged for an MIS graph $G$ is given (in $\hbar=1$ unit) by
\begin{equation}\label{HG}
\hat{H}({G}) =\sum_{(j,k) \in E(G)} U \hat{n}_{j} \hat{n}_{k} -\sum_{j \in V(G)} \left(\Delta \hat{n}_{j} -  \frac{\Omega}{2} \hat{\sigma}_{x}^{(j)} \right),
\end{equation}
where $U$ is the interaction between edged atoms, $\Omega$ and $\Delta$ are the Rabi frequency and detuning of Rydberg excitation, and $\hat{n}_{j}=(1-\hat{\sigma}_{z}^{(j)})/2$, $\hat{\sigma}_{x}^{(j)}$, $\hat{\sigma}_{z}^{(j)}$ are the excitation and Pauli operators defined for the ground ($\ket{0}$) and Rydberg ($\ket{1}$) states of the $j$-th atom. In the limit of $\Omega\rightarrow 0$, many-body ground states of $\hat H(G)$ correspond to the MIS solutions of $G$~\cite{Pichler2018,MHKim2022,Pichler2022}, because $U>0$ means the MIS problem's constraint that only one vertex can be in MIS for any two vertices on the same edge, satisfying the condition of the independent set, and $0<\Delta<U$ maximizes the number of vertices in the independent set. 

\begin{figure*}[thbp]
\centerline{\includegraphics[width=1\textwidth]{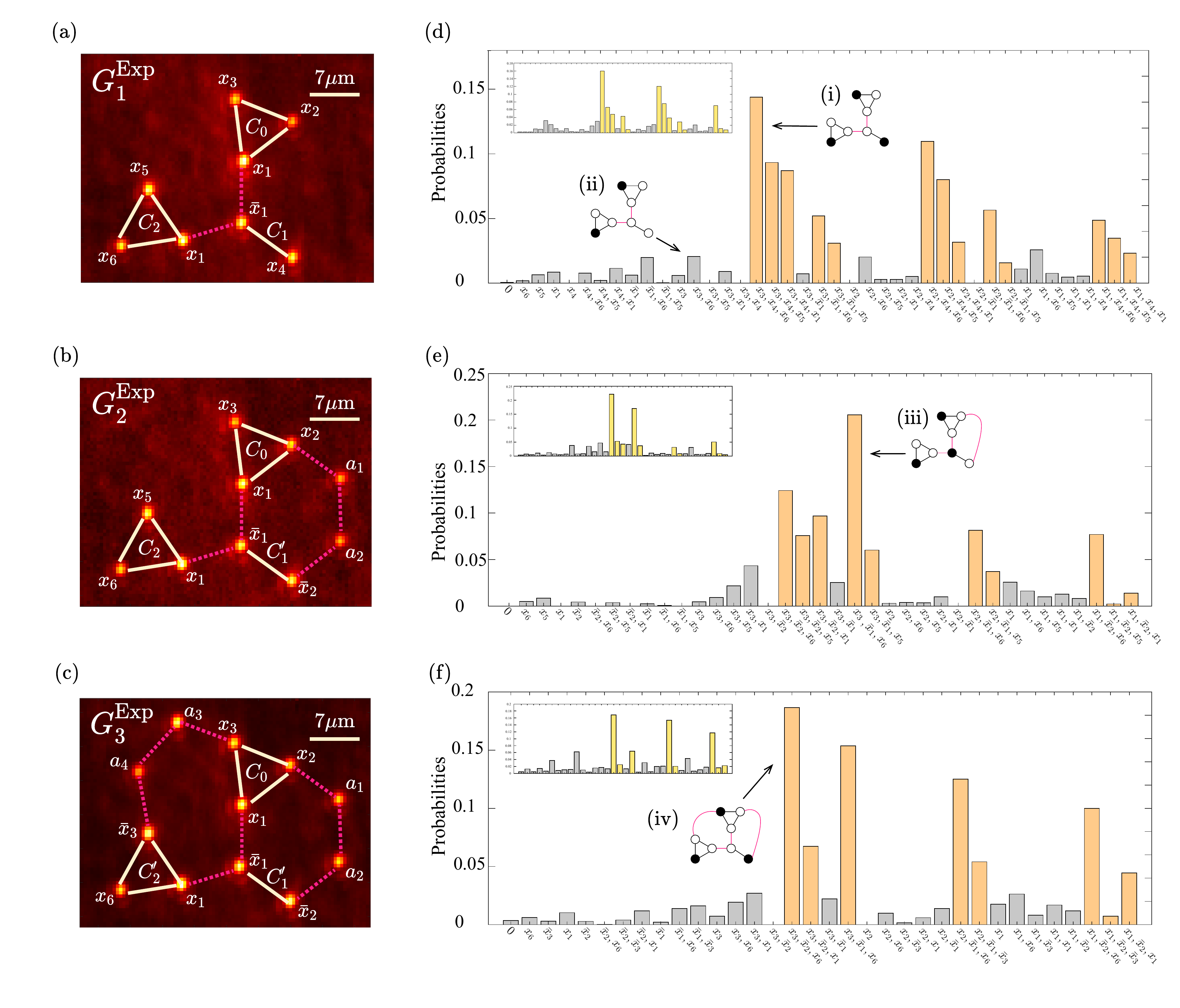}}
\caption{(a) Experimental MIS graph $G^{\rm Exp}_1$, (b) $G^{\rm Exp}_2$, and (c) $G^{\rm Exp}_3$ of literal atoms ($x_1,\cdots,x_6$) and quantum wire atoms ($a_1,\cdots,a_4$).  (d) Maximum likelihood probabilities of $G^{\rm Exp}_1$ experiments, where the $x$ axis denotes literal atoms in $\ket{1}$ in each binary configuration. (e) $G^{\rm Exp}_2$ experiments. (f) $G^{\rm Exp}_3$ experiments. For example, the peak (i) corresponds to $\ket{x_1x_2x_3;x_4;x_5x_6}=\ket{001;1;001}$, (ii) $\ket{001;0;001}$, (iii) $\ket{x_1x_2x_3;x_5x_6}=\ket{001;01}$, and (iv) $\ket{x_1x_2x_3;x_6}=\ket{001;1}$. Insets in (d-f) show numerical simulations with $\gamma=30$~$(2\pi)$~kHz of laser phase and dephasing noise taken into account.}
\label{Fig2}
\end{figure*}

The MIS graphs in Fig.~\ref{Fig1} are physically implemented with experimental graphs, $G^{\rm Exp}_1$, $G^{\rm Exp}_2$, and $G^{\rm Exp}_3$, in Figs.~\ref{Fig2}(a-c), where the ``normal'' edges (solid line edges) are between Rydberg blockaded pairs of atoms and long distance edges (dashed, inter-clause edges) are implemented with Rydberg quantum wires~\cite{MHKim2022,AndyByun2022,Qiu2020}. In Fig.~\ref{Fig2}(a) for $\Psi_1$, $x_1$ in $C_0$ and $C_2$ are edged to $\bar x_1$ in $C_1$, by placing the atom trios, $C_0$ (upper)-$C_1$ (lower right) and $C_1$ (lower right)-$C_2$ (lower left) closely, so that the atoms $x_1$ in $C_0$ (respectively, also in $C_2$) and $\bar x_1$ are at the distance $d$. In Fig.~\ref{Fig2}(b), the long edge between $x_2$ and $\bar x_2$ of $G^{\rm Exp}_2$ is implemented with a Rydberg quantum wire of two auxiliary atoms labeled by $\{a_1, a_2\}$. Also, in Fig.~\ref{Fig2}(c), the two long edges $x_2$-$\bar x_2$ and $x_3$-$\bar x_3$ of $G^{\rm Exp}_3$ are implemented respectively with two Rydberg quantum wires respectively with auxiliary atoms, $\{a_1, a_2\}$ and $\{a_3, a_4\}$, respectively. These wire atoms mediate the Rydberg blockade between two literal atoms with far distance~\cite{Weber2018,Qiu2020,MHKim2022}. All edged pairs of atoms including auxiliary atoms are at the same inter-atom distance, $d=7.0$~$\mu$m, smaller than the Rydberg blockade distance $d_B=10.0$~$\mu$m, and their two-dimensional positions are optimized for minimal unwanted (unedged) inter-atom interactions (see Supplementary for details). 

\begin{figure}[htbp]
\centerline{\includegraphics[width=0.5 \textwidth]{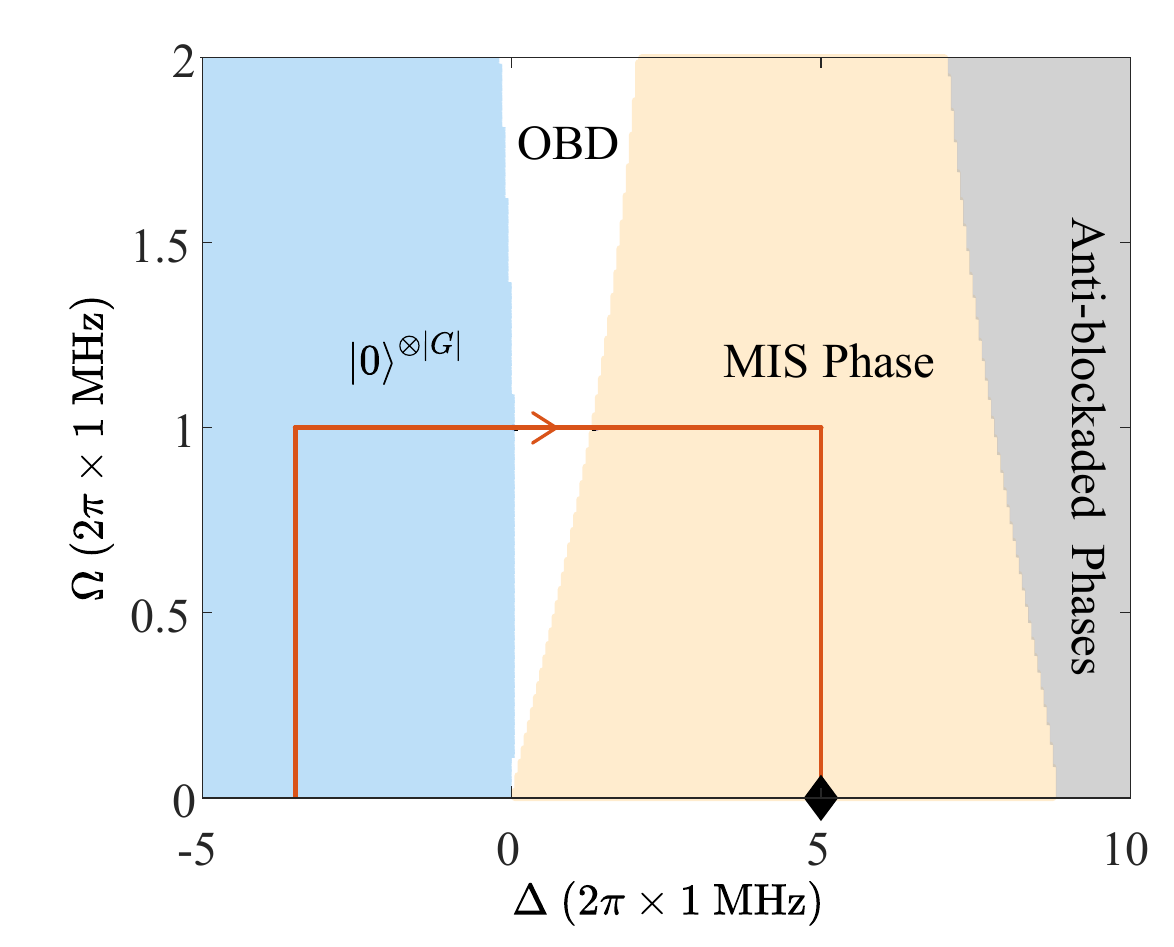}}
\caption{Phase diagram of $\hat{H}({G^{\rm Exp}_1})$ with the control path is shown with an arrow from the paramagnetic phase via the order-by-disorder (OBD) phase to the MIS phase. The phase diagrams of $\hat{H}(G^{\rm Exp}_{2})$ and $\hat{H}(G^{\rm Exp}_{3})$ are similar.}
\label{Fig3}
\end{figure}

Quantum computing of the 3-SAT problem starts with an experimental MIS graph $G$, chosen among $G^{\rm Exp}_{1,2,3}$, of atoms initially prepared in $\ket{0}^{\otimes |G|}$. We then turn on Rydberg excitation and adiabatically change the Hamiltonian from $\hat H(\Delta=-0.7\Delta_0, \Omega=0)$ for the paramagnetic phase to $\hat H(\Delta=\Delta_0, \Omega=0)$ for the MIS phase, along the control path denoted in the phase diagram~\cite{Scholl2021,Fey2019} in Fig.~\ref{Fig3}.
Experimental results are shown in Figs.~\ref{Fig2}(d-f) for $G^{\rm Exp}_1$, $G^{\rm Exp}_2$ and $G^{\rm Exp}_3$, respectively, where experimentally most-likely probabilities are plotted for all binary configurations of literal atoms with the $x$-axis denoting atoms in $\ket{1}$ only, and all anti-blockade atom configurations (of little probabilities) are omitted for the sake of presentation. The maximum-likelihood probability calculation~\cite{Teo2011} assumed state preparation and measurement (SPAM) errors~\cite{Sylvain2018}, $P\left(1\mid0\right)=3.9\%$ and $P\left(0\mid1\right)=7.9\%$ which are experimentally calibrated. Also the Rydberg quantum-wire compilation method~\cite{AndyByun2022} is used to impose the anti-ferromagnetic atom chain condition, $\ket{a_1a_2}$ and $\ket{a_3a_4} = \ket{01}$ or $\ket{10}$ among collected experimental data. In Figs.~\ref{Fig2}(d-f), orange bars are the 3-SAT solution states with one atom excited in each clause, while gray bars are non-solution states. For example, in Fig.~\ref{Fig2}(d) for $G^{\rm Exp}_1$, the maximal peak (i) of $\ket{x_1,x_2,x_3;\bar x_1 x_4;x_1,x_5,x_6}=\ket{001;01;001}$ is a 3-SAT solution and the peak (ii) of $\ket{001;00;001}$, in which two atoms in $C_1$ and $C_0$ are excited but none in $C_1$, is not a solution. Similarly,  most of the dominant peaks in Figs.~\ref{Fig2}(d-f) are verified to be 3-SAT solutions. For comparison, the insets of Figs.~\ref{Fig2}(d-f) show a numerical simulation of the same physical process traced with a Lindbladian equation taking into account experimental error sources such as the spontaneous decay rate ($\gamma/2\pi=30$~kHz) and laser phase noise (See Supplementary for details). The difference between the simulation and experiment is attributed to mainly the distance error between atoms and the laser beam center, which results in nonuniform Rabi frequencies of atoms in Eq.~\eqref{HG} (see Supplementary for details).

With the experimentally retrieved probabilities in Figs.~\ref{Fig2}(d-f), we can evaluate whether these three 3-SAT instances, $\Psi_{1,2,3}$, are solved correctly or not, i.e., whether the total probabilities of the MIS solution states are properly measured or not, for the satisfiability check of the 3-SAT instances. The probabilities of the orange bars in Fig.~\ref{Fig2}(d) for $G^{\rm Exp}_1$ are summed to be $81\%$, which, as a result, evaluates the satisfiability of $\Psi_1$ probabilistically very high. Likewise, the satisfiabilities of $\Psi_2$ and $\Psi_3$ are evaluated with probabilities of $78\%$ and $74\%$, obtained from the $G^{\rm Exp}_2$ and $G^{\rm Exp}_3$ experiments, respectively. 

\begin{figure}[t]
\centerline{\includegraphics[width=0.5\textwidth]{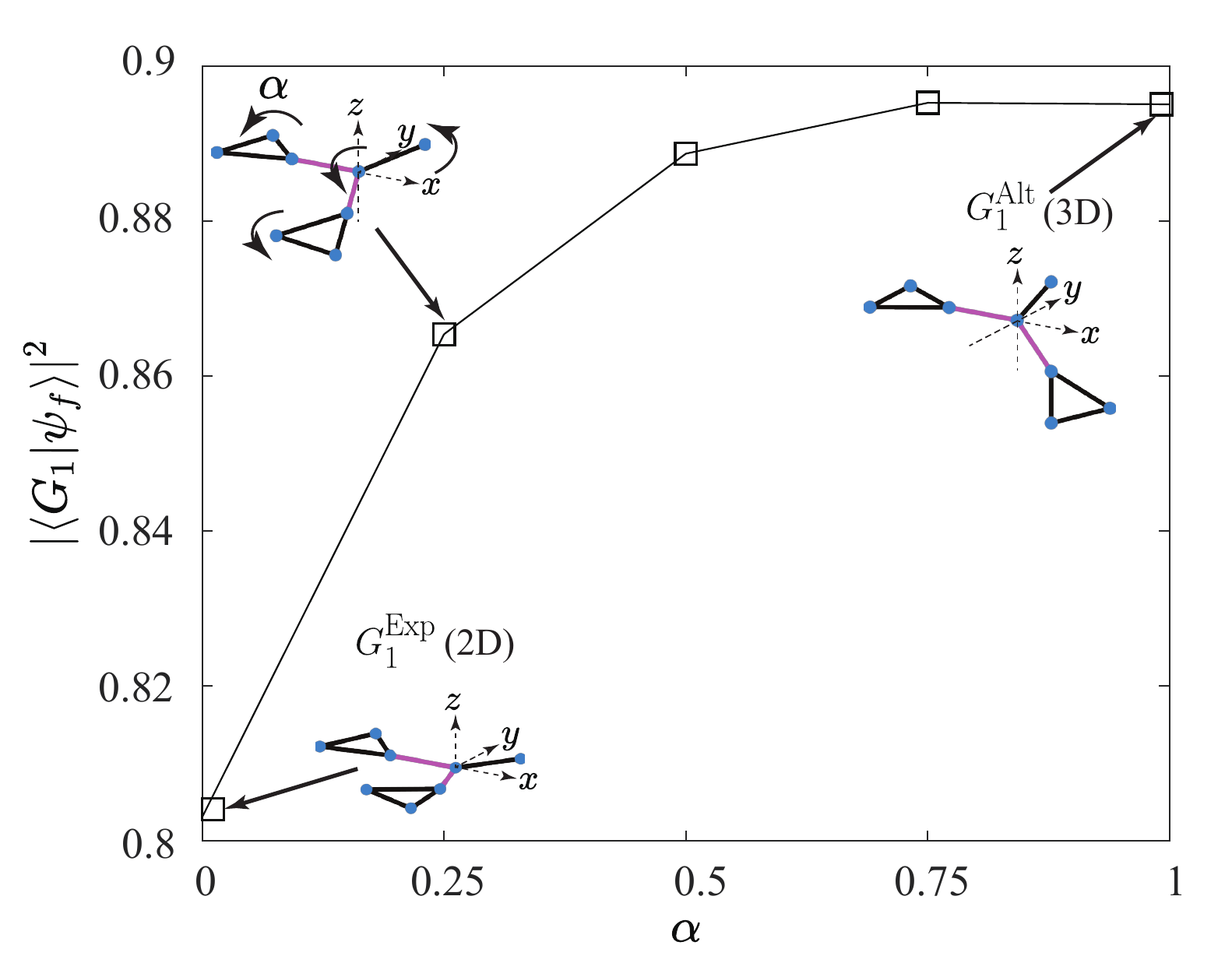}}
\caption{Ground state fidelity $\left\vert\braket{G_1}{\psi_f}\right\vert^2$ according to the structural deformation from $G_1^{\textup{Exp}}$ to $G_1^{\textup{Alt}}$ with respect to a normalized rotation angle $\alpha$.}
\label{Fig4}
\end{figure}

Three-dimensional Rydberg-atom graphs can improve the above experiments. The Hamiltonian $\hat H({G})$ in Eq.~\ref{HG} is an approximation, which neglects long-range interactions, so three-dimensional MIS graphs can be constructed for higher many-body ground state probabilities. In Fig.~\ref{Fig4}, we calculate the fidelity $|\langle{H(G_1^{\rm Exp})}|{\Psi_f}\rangle|^2$, where $|{H(G_1^{\rm Exp})}\rangle$ is the analytic many-body ground state of $ H(G_1^{\rm Exp})$ (see Supplementary for details) and $\ket{\Psi_f}$ is the numerically estimated final many-body state after the quasi-adiabatic evolution under our experimental condition (without decoherence taken into account). For the experimental graph $G_1^{\rm Exp}$ in 2D, the fidelity is estimated to be 81\%, due to the contribution of long-range residual Rydberg interactions among atoms spaced beyond the Rydberg blockade radius. In our atomic arrangements for $G_1^{\textup{Exp}}$, the average strength of the residual interactions is $\langle U_{\textup{res}}\rangle /2\pi = 0.64$~MHz, and their distribution is asymmetric. However, we can transform the structure of $G_1^{\textup{Exp}}$ to an alternative graph $G_1^{\textup{Alt}}$, which is more symmetric in geometry (see Supplementary for atomic positions). A structural transformation is conducted as in Fig.~\ref{Fig4}, by rotating the clauses $C_0$, $C_1$, and $C_2$ with respect to the edges from the literal atom $\bar x_1$ (central atom) respectively. These geometric changes are parameterized to a normalized rotation angle $\alpha$ (0 at $G_1^{\textup{Exp}}$, and 1 at $G_1^{\textup{Alt}}$). For the graph $G_1^{\textup{Alt}}$, the average residual interaction strength is reduced to $\langle U_{\textup{res}}\rangle / 2\pi = 0.40$~MHz. Then it is found that the ground state fidelity $|\langle{H(G_1^{\rm Exp})}|{\Psi_f}\rangle |^2$ after the same quasi-adiabatic evolution is improved to 90\%. So three-dimensional atom allocations have more degrees of freedom for better experimental performances. Fig.~\ref{Fig4} shows the improvement of the ground state fidelity $\left\vert\braket{G_1}{\psi_f}\right\vert^2$ during the structural transformation.

While all MIS graphs are in principle implementable in the three-dimensional space with quantum wires~\cite{MHKim2022}, it is worthwhile to discuss the scaling issue of the Rydberg-atom approach to the 3-SAT problem. First, we estimate the number of atoms, $N_A$, necessary for general 3-SAT instances. For a Boolean expression that has $N_C$ clauses, the lower bound is $N_A \gtrsim 3N_C$. The upper bound is the case of maximal literal-negation pairs, so the corresponding MIS graph has maximal inter-clause interactions. Physical implementation requires auxiliary atoms using either `crossing lattice' scheme~\cite{Pichler2022_1} or `quantum wire' scheme~\cite{MHKim2022}, both of which are recently suggested and experimentally demonstrated. In the `crossing lattice' scheme, each vertex is transformed to an atom chain on a 2D surface, and the interactions between the vertices are implemented with `crossing gadgets' of at most 8 atoms. So, the upper bound of $N_A$ is estimated to be $36N_C^2$ for an $3N_C$-vertex non-unit disk graph. In the `quantum wire' scheme, the required number of total atoms is numerically estimated with the Erdos-Renyi model for $3N_C$-vertex graphs of maximum degree 6 in 3D~\cite{Dalyac2022},  to be linear to $N_C$.

Next we estimate an experimental time budget to access a large-scale 3-SAT problem. The probability to successfully obtain the solutions of the 3-SAT problem, after $M$ experimental repetitions, is given by
\begin{equation}\label{P_s}
P_{s}(p,M)=\sum_{j=1}^{M} (1-p)^{j-1} p = 1-(1-p)^M,
\end{equation}
where $p$ is the ground-state probability of the corresponding MIS graph of $N_A$ atoms. With the experimental scaling $p\sim 1.04^{-N_A}$ of a state-of-the-art experimental platform~\cite{Pichler2022}, the ground-state probability of an $N_A=400$ MIS graph is estimated to be $p(N_A=400)\sim 10^{-7}$. The required number of repetitions to achieve $P_s>20$\%, for example, is given by $M> \log 0.8/\log (1-p)$, which estimates about $M\sim 10^6$ experimental repetitions. So, an $N_A=400$ MIS graph experiment, which can serve 3-SAT instances with approximately 12$\sim$140 clauses, would take one week in the typical repetition rate of 2$\sim$3~Hz of the current experimental platforms.

In summary, Rydberg atom interactions can be used to program Boolean expressions of the 3-SAT problem and determine their satisfiabilities using the presented 3-SAT quantum algorithm which combines the reduction algorithm from SAT (or 3-SAT) to MIS and the ground-state searching algorithm of Rydberg atom graphs. This result indicates that all other NP problems are also programmable (by Cook-Levin theorem) with Rydberg atoms, either by using the 3-SAT quantum algorithm as a subroutine or directly formulating a quantum algorithm for Rydberg atoms. While it is difficult to conclude the technical limits of this still-nascent Rydberg atom technology, this work is clearly paving a new route to quantum computing developments.

This research is supported by Samsung Science and Technology Foundation (SSTF-BA1301-52).

\end{document}